\documentclass[showpacs,amsmath,amssymb,aps]{revtex4}
\usepackage{graphicx}
\usepackage{amsthm}
\usepackage{epsfig}
\def\d{\operatorname{d}}\def\<{\langle}\def\>{\rangle}
\def\Tr{\operatorname{Tr}}
\def\set#1{{\sf #1}}
\def\grp#1{{\mathbf #1}}
\def\Reals{\mathbb R}
\def\spc#1{\mathcal{#1}}
\def\Proof{\medskip\par\noindent{\bf Proof. }}
\def\qed{$\,\blacksquare$\par}

\newtheorem{prop}{Proposition}
\begin{document}
\title{Joint estimation of real squeezing and displacement}
\author{G.~Chiribella} \email{chiribella@unipv.it}
\author{G.~M.~D'Ariano} \email{dariano@unipv.it} \altaffiliation[Also
at ]{Center for Photonic Communication and Computing, Department of
  Electrical and Computer Engineering, Northwestern University,
  Evanston, IL 60208} \affiliation{QUIT Quantum Information Theory
  Group, 
Dipartimento di Fisica ``A. Volta'', via Bassi 6, I-27100
  Pavia, Italy}\homepage{http://www.qubit.it}
 \author{M.~F.~Sacchi} \email{msacchi@unipv.it}
\affiliation{QUIT Quantum Information Theory Group, CNR - Istituto
  Nazionale per la Fisica dell Materia, via Bassi 6, I-27100 Pavia,
  Italy} 
\homepage{http://www.qubit.it}

\begin{abstract}  
  We study the problem of joint estimation of real squeezing and
  amplitude of the radiation field, deriving the measurement that
  maximizes the probability density of detecting the true value of the
  unknown parameters. More generally, we provide a solution for the
  problem of estimating the unknown unitary action of a nonunimodular
  group in the maximum likelihood approach.  Remarkably, in this case
  the optimal measurements do not coincide with the so called
  square-root measurements. In the case of squeezing and displacement
  we analyze in detail the sensitivity of estimation for coherent
  states and displaced squeezed states, deriving the asymptotic
  relation between the uncertainties in the joint estimation and the
  corresponding uncertainties in the optimal separate measurements of
  squeezing and displacement.  A two-mode setup is also analyzed,
  showing how entanglement between optical modes can be used to approximate  perfect
  estimation.
\end{abstract}


\maketitle 
\section{Introduction}
Squeezing and displacement are the basic operations of continuous
variables quantum information \cite{BP}, and are easily performed, the
former by parametric amplifiers, the latter by lasers and linear
optics. Squeezing in the two-mode setup is, for example, the tool to
generate entanglement in the Braunstein-Kimble teleportation scheme
\cite{BK}. The combined use of real squeezing and displacement allows
one to encode efficiently classical information in quantum channels
using homodyne detection at the receiver \cite{CavesRev}. In a quantum
communication scenario where a coherent signal is sent through a
non-linear medium and undergoes an amplification process, the joint
estimation of displacement and squeezing provides a twofold
information: the amplitude modulation of the state and a property of
the communication channel itself. This may be useful in a
communication scheme designed to be robust to photon loss.

In this paper we consider the problem of jointly estimating real
amplitude and real squeezing of the radiation field. In a tomographic
setup, where a large number of equally prepared copies is available,
the maximum likelihood method turns out to be very efficient in
estimating the parameters that characterize the state of the quantum
system \cite{MlReconstruction,MlMethods}. In this approach, first one
fixes a set of single-copy measurements (typically homodyne
measurements at some random phase) and then looks for the estimate
that maximizes the probability (density) of producing the observed
data.  However, the tomographic approach is not suitable to the case
when only a small number of copies is available, and one needs to use
the limited resources at disposal more efficiently. It becomes then
important to optimize not only the posterior processing of the
experimental data, but also the choice of the measurement that is used
to extract these data from the system.  The most natural framework to
deal with this situation is quantum estimation theory
\cite{Helstrom,Holevo}, where the concept of positive operator valued
measure (POVM) provides a tool to describe at the same time both
measurement and data processing. The maximum likelihood approach in
quantum estimation theory \cite{MlHolevo} then corresponds to seek the
measurement that maximizes the probability (density) that the
estimated value of the parameters coincides with the true value.

Joint estimation of squeezing and displacement is equivalent to infer
an unknown transformation of a group---in the present case the affine
group. This is an example of a frequent situation in quantum estimation, especially
in communication problems, where a set of signal states is generated
from a fixed input state by the action of a group.  Consider, for
example, the case of phase estimation for high-sensitivity
interferometry and optimal clocks \cite{DMS,HolevoPhase, OptClocks},
the estimation of rotation for the optimal alignment of reference
frames \cite{Refframe}, and the estimation of displacement of the
radiation field for the detection of a coherent signal in Gaussian
noise \cite{YuenLax}. In these cases, the symmetry of the problem
provides a physical insight that allows one to simplify the search for
efficient estimation strategies, and the concept of \emph{covariant}
measurement \cite{Holevo} becomes crucial for optimization.  The use
of maximum likelihood approach in the covariant setting has been shown
to be particularly successful in obtaining explicitly the optimal
measurements and in understanding the fundamental mechanism that leads
to the ultimate sensitivity of quantum measurements \cite{CovLik,
  DeGiorgi}. Group theoretical tools, such as equivalent
representations \cite{CovLik} and multiplicity spaces \cite{DeGiorgi},
far from being abstract technicalities, are the main ingredients to
achieve the ultimate quantum limits for sensitivity
\cite{Refframe,EntEstimation}.  Moreover, in a large number of
situations, the maximization of the likelihood provides measurements
that are optimal also according to a wide class of different figures
of merit \cite{EntEstimation}.

From a group theoretical point of view, the action of real squeezing
and displacement on the wavefunctions provides a unitary
representation of the affine group ``$ax +b$'' of dilations and
translations on the real line. The structure of the affine group
underlies the theory of wavelets \cite{wavelets}, and has been
recently used in the characterization of $SU(1,1)$ coherent states
\cite{HayashiSaka,Bertrand}, and in the study of oscillators in a Morse
potential \cite{Morse}.

The affine group is particularly
interesting, since it is the paradigmatic example of a nonunimodular
group, namely a group where the left-invariant Haar measure is
different from the right-invariant one. This leads to orthogonality
relations that---differently from the usual Schur lemmas---involve a
positive unbounded operator, firstly introduced by Duflo, Moore, and
Carey (DMC) in Ref. \cite{Duflo}. As we will see in this paper, the
nonunimodularity of the group has some amazing consequences in the
estimation problem. For example, the so-called square-root
measurements \cite{SquareRoot}, that are commonly considered in
quantum communication and cryptography, do not coincide with the
maximum-likelihood measurements, the latter providing a higher
probability density of correct estimation. Another bizarre feature is
that for maximum likelihood measurements, the most likely value in the
probability distribution can be different from the true one. While for
unimodular groups this feature never happens, for nonunimodular groups
it is unavoidable, and a suitable choice of the input states is needed
to reduce the discrepancy between the true value and the most likely
one.

The paper is organized as follows. In order to set the optimal joint
estimation of real squeezing and displacement in the general
estimation method, first we derive in Sec. \ref{MlEst} the
optimal measurement for the problem of estimating the unknown unitary action
of a nonunimodular group.  Then, as a special example, the general
results will be used to optimize the joint estimation of squeezing and
displacement in Sec. \ref{JointEst}. The efficiency of coherent states and displaced squeezed states is analyzed in detail, and the asymptotic relation between the uncertainties in the joint estimation and the uncertainties in the optimal separate measuremnts of squeezing and displacement is derived.  The conclusions are summarized
in Sec. IV. The explicit derivation of group average over the affine
group is given in the Appendix. 
\section{Maximum likelihood estimation for a nonunimodular group}\label{MlEst}
\subsection{Background and generalities}
Suppose that a fixed input state $\rho$, corresponding to a density operator
in the Hilbert space $\spc H$, is transformed by the unitary
representation $\{U_g ~|~g \in \grp G\}$ of the group $\grp G$, so that it generates the family of signal states
\begin{equation}\label{Orbit}
\mathcal{O} = \{U_g \rho U_g^{\dag}~|~ g \in \grp G\}~.
\end{equation}
The typical quantum estimation problem is then to find the measurement
that gives the best estimate for the unknown transformation $g$
according to some optimality criterion.  Usually the criterion is
given by a cost function $c(\hat g, g)$, which quantifies the cost of
estimating $\hat g$ when the true value is $g$, and enjoys the
invariance property $c(h
\hat g,h g)=c(\hat g, g) \quad \forall h, \hat g,g \in \grp G$.  For
example, the maximum likelihood criterion corresponds to the delta
cost-function $c(\hat g, g)= -\delta (g^{-1} \hat g)$ (loosely
speaking, there is an infinite gain if the estimate coincides with the
true value, and no gain otherwise).  Once a cost function is fixed,
one can choose two possible approaches to optimization, namely the
Bayes approach and the minimax. In the Bayes approach, one assumes
some prior distribution of the unknown parameters, and then minimizes
the average over the true values of the expected cost $c(g)= \int \d
\hat g~ c(\hat g, g) p(\hat g|g)$, where $p(\hat g|g)$ is the
conditional probability density of estimating $\hat g$ when the true
value is $g$.  In the minimax approach, one looks instead for the measurement
that minimizes the supremum of the expected cost over all possible true
values of the unknown parameters.

An important class of estimation strategies is given by the
\emph{covariant} measurements, that are described by POVMs of the form
\cite{Holevo}
\begin{equation}\label{CovPovm}
M(g)= U_g \xi U_g^{\dag}~,
\end{equation}
where $\xi \ge 0$ is an operator satisfying the normalization condition
\begin{equation}
\int_{\grp G} \d_L g~ U_g \xi U_g^{\dag}= \openone~,\label{nc}
\end{equation} 
$\d_L g$ denoting the left-invariant Haar measure on the group, namely
\mbox{$\d_L (h g) = \d_L g \quad \forall h,g \in \grp G$}.  Due to the
symmetry of the set of states (\ref{Orbit}), covariant measurements play a fundamental role in
the search of the optimal estimation. For compact groups, the
following proposition holds:
\begin{prop}[Holevo \cite{Holevo}]\label{CovMinimax} 
    For compact groups, the search for the optimal measurement in the minimax
  approach can be restricted without loss of generality to the class
  of covariant measurements.
\end{prop}
Moreover, for compact groups the optimality of covariant
measurements holds also in the Bayes approach \cite{Holevo}, if the prior
distribution is chosen to be the normalized Haar measure on the group,
i.e. the measure $\d g$ such that $\d g = \d (h g) =\d (g h) \quad
\forall g ,h \in \grp G$ and $\int _{\grp G}~ \d g=1$.

For non-compact groups, such as the affine group involved in the joint
estimation of squeezing and displacement, the situation is more
involved. Of course, in the Bayes approach it is no longer possible to
choose the uniform Haar measure as prior distribution, since it is not
normalizable.  However, in the minimax the optimality of covariant
measurements still holds, even though in this case the proof becomes
rather technical \cite{OzawaUnpublished}. In this paper we will adopt the minimax approach to deal with noncompact groups, and this will allow us to restrict the optimization to covariant POVMs.  
\subsection{Nonunimodular groups}
As we will see in Section III, the action of real squeezing and
displacement on one mode of the radiation field yields a
representation of the affine group ``$ax +b$'' of dilations and
translations on the real line. This group is clearly non-compact, and, moreover, it is nonunimodular, namely the left-invariant Haar measure $\d_L g$
(with $\d_L (hg) = \d_L g \quad \forall h,g \in \grp G$) does not
coincide with the right-invariant one $\d_R g$ (with $\d_R (gh)=\d_R g
\quad \forall h,g \in \grp G$). Therefore, to face the estimation
problem with the affine group, we need some results about
representation theory and orthogonality relations for nonunimodular
groups (for an introduction to these topics, see for example
\cite{GrossmannMorlet}).

Let $\{U_g ~|~ g \in \grp G\}$ be a unitary representation of a locally
compact group $\grp G$. In the following, we will make two assumptions
on the representation $\{U_g\}$ that are tailored on the concrete problem of estimating real squeezing and displacement.

\emph{First assumption: discrete Clebsch-Gordan series.} We require
the representation $\{U_g\}$ to be a direct sum of irreducible
representations (\emph{irreps}), namely that its Clebsch-Gordan series
is discrete.  In this case, there is a decomposition of the Hilbert
space $ \spc H$ as
\begin{equation}\label{SpaceDecomp}
\spc H = \bigoplus_{\mu \in \set S}~ \spc H_{\mu} \otimes \spc M_{\mu}~,
\end{equation}
such that
\begin{equation}\label{RepDecomp}
U_g = \bigoplus_{\mu \in \set S}~ U^{\mu}_g \otimes \openone_{\spc
M_{\mu}}~, \end{equation} where $\{U_g^{\mu}\}$ is an irreducible
representation acting on the Hilbert space $\spc H_{\mu}$ and
$\openone_{\spc M_{\mu}}$ is the identity in the space $\spc
M_{\mu}$. The Hilbert spaces $\spc H_{\mu}$ and $\spc M_{\mu}$ are called
\emph{representation} and \emph{multiplicity} spaces,
 respectively, and the index
$\mu$ labels the inequivalent irreps that appear in the Clebsch-Gordan
series $\set S$ of the representation $\{U_g\}$.

\emph{Second assumption: square-summable irreps.} We require each
irreducible representation $\{U^{\mu}_g\}$ in Eq. (\ref{RepDecomp}) to
be square-summable. This means that there is at least one non-zero
vector $|\psi_{\mu}\> \in \spc H_{\mu}$ such that
\begin{equation}
\int_{\grp G} \d_L g~ |\<\psi_{\mu}| U_g^{\mu} |\psi_{\mu}\>|^2 < \infty~. 
\end{equation}
Vectors such that the above integral converges are called
\emph{admissible}. It is possible to show \cite{GrossmannMorlet} that,
if a representation is square-summable, then the set of admissible
vectors is dense in the Hilbert space.

Let us consider now the group average of an operator $A$, defined as
\begin{equation}\label{DefAve}
\<A\>_{\grp G}= \int_{\grp G} \d_L g~ U_g A U_g^{\dag}~.
\end{equation}
In general, the average may not converge for any operator (for
example, it diverges for $A= \openone$). In analogy with admissible
vectors, we say that $A$ is an \emph{admissible operator} if the group
average in Eq. (\ref{DefAve}) converges in the weak operator sense.
In such a case, one can prove that $\<A\>_{\grp G}$ is given by
\begin{equation}\label{AveOp}
\<A\>_{\grp G}=  \bigoplus_{\mu \in \set S} \openone_{\spc H_{\mu}} \otimes \Tr_{\spc H_{\mu}} [D_{\mu} \otimes \openone_{\spc M_{\mu}}~ A]~,  
\end{equation}
where $D_{\mu}$ is a positive self-adjoint operator acting on the
representation space $\spc H_{\mu}$. The operator $D_{\mu}$ has been
firstly introduced by Duflo, Moore, and Carey (DMC) \cite{Duflo}, and
is the characteristic feature of nonunimodular groups. In fact, if the
group $\grp G$ is unimodular---i.e. if the left- and right-invariant
measures coincide---than the DMC operator is simply a multiple of the
identity, and the formula (\ref{AveOp}) for the group average is
equivalent to the ordinary Schur lemmas. Contrarily, if the group
$\grp G$ is nonunimodular, the DMC operator is a positive unbounded
operator, and its presence modifies the orthogonality relations
dramatically, with remarkable consequences in the estimation of an unknown group transformation.
 
The admissibility of vectors and operators has a simple
characterization in terms of the DMC operator. As regards vectors, the
set of admissible vectors for the irrep $\{U_g^{\mu}\}$ can be
characterized as the domain of $\sqrt{D_{\mu}}$. For unimodular
groups, since $D_{\mu}$ is proportional to the identity, the set of
admissible vectors is the whole representation space $\spc H_{\mu}$,
while for nonunimodular groups the admissible vectors form a dense
subset of $\spc H_{\mu}$.  As regards operators, an operator $A$ is
admissible if and only if the partial traces $\Tr_{\spc H_{\mu}}
[D_{\mu} \otimes \openone_{\spc M_{\mu}}A]$ are not diverging. For
example, if $|\psi_{\mu}\> \in \spc H_{\mu}$ is an admissible vector
and $O_{\mu}$ is any operator acting on $\spc M_{\mu}$, then the
operator $A_{\mu}=|\psi_{\mu}\>\<\psi_{\mu}| \otimes O_{\mu}$ is
admissible, and its group average is $\<A_{\mu}\>_{\grp G}=
\<\psi_{\mu}|D_{\mu}|\psi_{\mu}\>~ \openone_{\mu} \otimes O_{\mu}$.

\subsection{Maximum likelihood measurements}
In order to find the best estimate for the signal states (\ref{Orbit})
according to the maximum likelihood criterion in the minimax approach,
we consider now a covariant measurement, described by a POVM $M(g)$ as
in Eq. (\ref{CovPovm}). The normalization condition (\ref{nc}) can be
rewritten as
\begin{equation}\label{Norm}
\int_{\grp G}~ \d_L g~ M(g)= \<\xi\>_{\grp G}= \openone~.
\end{equation}   

Using Eq. (\ref{AveOp}) for evaluating the group average
$\<\xi\>_{\grp G}$, this condition becomes
\begin{equation}\label{TrXi}
\Tr_{\spc H_{\mu}} [ D_{\mu} \otimes \openone_{\spc M_{\mu}}~\xi]= \openone_{\spc M_{\mu}}~.
\end{equation}  

According to the maximum likelihood approach, we need to find the
covariant measurement that maximizes the probability density that the
estimated transformation coincides with the true one, i.e.
$\mathcal{L}= p(g|g)= \Tr[M(g) U_g \rho U_g^{\dag}]=\Tr[\xi \rho]$.
For a pure input state $\rho=|\psi\>\<\psi|$, the maximization of the
likelihood over all possible operators $\xi \ge 0$ satisfying the
constraints (\ref{TrXi}) follows in a simple way by a repeated use of
Schwartz inequality. In fact, the input state $|\psi\>$ can be written
in the decomposition (\ref{SpaceDecomp}) as
\begin{equation}\label{StateDecomp}
|\psi\>= \bigoplus_{\mu \in \set S}~ c_{\mu} ~|\Psi_{\mu}\>\!\>~,
\end{equation}     
where each $|\Psi_{\mu}\>\!\> \in \spc H_{\mu} \otimes \spc M_{\mu}$
is a bipartite state.  From Schwartz inequality, we have
\begin{equation}
\mathcal{L}\le \sum_{\mu,\nu}~\left| c_{\mu}^* c_{\nu}~ \<\!\<\Psi_{\mu}|\xi|\Psi_{\nu}\>\!\>\right| \le \left( \sum_{\mu} |c_{\mu}| \sqrt{\<\!\<\Psi_{\mu}|\xi|\Psi_{\mu}\>\!\>}\right)^2  \end{equation}
At this point, we assume that each bipartite state
$|\Psi_{\mu}\>\!\>$ is in the domain of the operator $D_{\mu}^{-1/2}
\otimes \openone_{\spc M_{\mu}}$. This assumption is not 
restrictive, since the domain of a self-adjoint operator is dense in
the Hilbert space. In this way, it is possible to write
\mbox{$|\Psi_{\mu}\>\!\>= D_{\mu}^{1/2} D_{\mu}^{-1/2}\otimes 
\openone_{\spc M_{\mu}} |\Psi_{\mu}\>\!\>$} and to exploit the Schmidt
decomposition of the (non-normalized) vector $D_{\mu}^{-1/2} \otimes
\openone_{\spc M_{\mu}}~|\Psi_{\mu}\>\!\>$. In other words, we can
write
\begin{equation}\label{BipartiteMu}
 |\Psi_{\mu}\>\!\>= \sum_{m=1}^{r_{\mu}}~
 \sqrt{\lambda_m^{\mu}}~\sqrt{D_{\mu}} \otimes \openone_{\spc M_{\mu}}
 |\tilde\psi_m^{\mu}\>|\tilde\phi_m^{\mu}\>~,
\end{equation}
where $r_{\mu}$ is the Schmidt rank, $\lambda_m^{\mu} \ge 0$ are
Schmidt coefficients such that $\sum_{m=1}^{r_\mu } \lambda_m^{\mu} =
\<\!\<\Psi_{\mu}|D_{\mu}^{-1} \otimes \openone_{\spc
  M_{\mu}}|\Psi_{\mu}\>\!\>$, and $|\tilde \psi^{\mu}_m\>,
|\tilde\psi^{\mu}_m\>$ are the elements of two orthonormal bases for
$\spc H_{\mu}$ and $\spc M_{\mu}$, respectively.  The form
(\ref{BipartiteMu}) is very convenient for optimization, in fact we
can use again Schwartz inequality and obtain
\begin{equation}
\begin{split}
\<\!\<\Psi_{\mu}|\xi|\Psi_{\mu}\>\!\> &\le \left( \sum_m \sqrt{\lambda_m^{\mu} ~ \<\tilde\psi_m^{\mu}|\<\tilde\phi_m^{\mu}| \sqrt{D_{\mu}} \otimes \openone_{\spc M_{\mu}} \xi \sqrt{D_{\mu}} \otimes \openone_{\spc M_{\mu}}|\tilde\psi_m^{\mu}\>|\tilde\phi_m^{\mu}\>} \right)^2\\
\end{split}
\end{equation}
Finally, we have
\begin{eqnarray}
\<\tilde\psi_m^{\mu}| \<\tilde\phi_m^{\mu}| \sqrt{D_{\mu}}\otimes
\openone_{\spc M_{\mu}} \xi \sqrt{D_{\mu}} \otimes \openone_{\spc
  M_{\mu}} |\tilde\psi_m^{\mu}\>|\tilde\phi_m^{\mu}\> 
\le \<\tilde\phi_m^{\mu}|~\Tr_{\spc H_{\mu}}[D_{\mu} \otimes 
\openone_{\spc M_{\mu}}\,\xi ]~|\tilde\phi_m^{\mu}\>=1~,
\end{eqnarray} 
the last equality following from the normalization constraint (\ref{TrXi}).
Therefore, the previous chain of inequalities proves the upper bound
\begin{equation}\label{OptLik}
\mathcal{L} \le \left(\sum_{\mu \in \set S} |c_{\mu}|~ \sum_{m=1}^{r_{\mu}} \sqrt{\lambda_m^{\mu}}\right)^2=\mathcal{L}^{opt}~,  
\end{equation}
that holds for any covariant POVM.  On the other hand, it is immediate
to check that the bound is achieved by the covariant POVM given by
$\xi =|\eta\>\<\eta|$ with
\begin{equation}\label{OptEta}
|\eta\>= \bigoplus_{\mu \in \set S} e^{i \arg (c_{\mu})} ~\sum_{m=1}^{r_{\mu}} D_{\mu}^{-1/2} \otimes \openone_{\spc M_{\mu}}~ |\tilde\psi_m^{\mu}\>|\tilde\phi_m^{\mu}\>~,  
\end{equation}
$\arg (z)$ denoting the argument of a complex number, i.e. $z=|z|
e^{i\arg (z)}$. The normalization of such a POVM follows from Eq.
(\ref{TrXi}), and one has 
\begin{eqnarray}
\int \d _L g U_g \xi U^\dag _g =\bigoplus _\mu \openone _{\spc
  H_{\mu}} \otimes \sum _{m=1}^{r_\mu } |\tilde \phi ^\mu _m \rangle
\langle \tilde \phi _m ^\mu |\;,\label{}
\end{eqnarray}
namely, the POVM is complete in the subspace spanned by the orbit of
$| \psi \rangle $, and can be trivially completed to the whole Hilbert
space without affecting the probability distribution. Notice that, if
the group $\grp G$ is unimodular, namely $D_{\mu}= \openone_{\spc
  H_{\mu}} /d_{\mu}$ for some positive constant $d_{\mu}$, then we
correctly retrieve the results of Ref.  \cite{DeGiorgi} about
maximum-likelihood measurements.
\par 
\emph{The case of $\{U_g\}$ being a direct sum of inequivalent irreps.} The expression for the optimal covariant POVM can be further
simplified in the case when all the multiplicity spaces $\spc M_{\mu}$
are one-dimensional, i.e. when the representation $\{U_g\}$ is a
direct sum of inequivalent irreps. In this case, we can decompose the
input state as $|\psi\>= \bigoplus_{\mu}~c_{\mu} |\psi_{\mu}\>$ (as in
Eq. (\ref{StateDecomp}), but without the need of introducing bipartite
states), and now the decomposition (\ref{BipartiteMu}) becomes
trivial, namely
\begin{equation} 
|\psi_{\mu}\>= \sqrt{\lambda_{\mu}}~ \sqrt{D_{\mu}} |\tilde \psi_{\mu}\>~,
\end{equation} 
where $\lambda_{\mu}=  \<\psi_{\mu}| D_{\mu}^{-1} |\psi_{\mu}\>$ and 
\begin{equation}
|\tilde \psi_{\mu}\>= \frac{D_{\mu}^{-1/2}|\psi_{\mu}\>}{||
D_{\mu}^{-1/2} |\psi_{\mu}\>||}~.
\end{equation}   
Therefore, Eq. (\ref{OptEta}) for the optimal POVM becomes
\begin{equation}\label{OptEtaIneq}
|\eta\>= \bigoplus_{\mu\in \set S}~e^{i \arg(c_{\mu})}~
\frac{D_{\mu}^{-1} |\psi_{\mu}\>}{\sqrt{\<\psi_{\mu}| D_{\mu}^{-1}
    |\psi_{\mu}\>}}~=~ \bigoplus_{\mu \in \set S}~ \frac{D_{\mu}^{-1}
  |\psi\>}{\sqrt{\<\psi| D_{\mu}^{-1} |\psi\>}}~,  
\end{equation} 
and the corresponding optimal likelihood is given by
\begin{equation}\label{OptLikIneq}
\mathcal{L}^{opt}= |\<\eta|\psi\>|^2=\left( \sum_{\mu 
\in \set S}  \sqrt{\<\psi | D_{\mu}^{-1}|\psi \>} \right)^2~. 
\end{equation} 
\subsection{Remarks}
\emph{Remark 1: Square-root measurements.} A possible strategy to
estimate an unknown quantum state, randomly drawn from a given family,
is given by the so-called square-root measurements (SRM), firstly introduced
by Hausladen and Wootters
\cite{SquareRoot}.  In the case of pure states with a group symmetry as in Eq. (\ref{Orbit}),
there is an important connection between SRM and maximum-likelihood
measurements.  For example, if the group $\grp G$ is a discrete group
of phase shifts, it has been proved in
Refs. \cite{SRMBan,SRMSasaki,SRMMit}, that SRM minimize the
probability of error in estimating the unknown state,
and, equivalently, they maximize the likelihood, i.e. the
probability of correct estimation. More generally, the optimality of
the SRM in the maximum likelihood approach has been proved in
Ref. \cite{DeGiorgi} for a large class of groups, including all finite
groups, all compact groups, and unimodular noncompact groups, such as
the Weyl-Heisenberg group of displacements.  However, as we will see in the following, the case of nonunimodular groups represents an exception to the fact that SRM are optimal for the maximum likelihood criterion in the presence of a physical symmetry. In fact, the
SRM for the estimation of a group transformation
acting on a fixed state $\rho$ is given by the POVM $M_{sq}
(g)=F^{-1/2}~ U_g \rho U_g^{\dag}~ F^{-1/2}$, where
\begin{equation}\label{FrameOp}
F= \int_{\grp G} \d_L g~ U_g \rho U_g^{\dag}
\end{equation}      
(the POVM $M_{sq}(g)$ is obviously normalized with respect to the
left-invariant measure $\d_L g$).  The comparison with the
maximum-likelihood measurements of the previous section is
particularly simple in the case of group representations $\{U_g\}$
that are direct sum of inequivalent irreps.  In fact, for a pure state
$\rho =|\psi\>\<\psi|$ with $|\psi\>= \bigoplus_{\mu}~ c_{\mu}
|\psi_{\mu}\>$, the integral (\ref{FrameOp}) is easily calculated by
using Eq. (\ref{AveOp}), namely $F= \bigoplus_{\mu} |c_{\mu}|^2
\<\psi_{\mu}|D_{\mu}|\psi_{\mu}\>~ \openone_{\spc H_{\mu}}$. Notice
here that the square-root measurement can be defined only if $|\psi
_\mu \rangle $ is in the domain of $D_\mu ^{1/2}$. Therefore, the
square-root measurement is given by the covariant POVM $M_{sq}(g)= U_g
|\eta_{sq}\>\<\eta_{sq}| U_g^{\dag}$, where
\begin{equation}
|\eta_{sq}\>= \bigoplus_{\mu} e^{i \arg(c_{\mu})}~ 
\frac{|\psi_{\mu}\>}{\sqrt{\<\psi_{\mu}|D_{\mu}|\psi_{\mu}\>}}~.
\end{equation} 
This covariant POVM is different from the optimal one given in
(\ref{OptEtaIneq}), and does not achieve the optimal value
(\ref{OptLikIneq}) for the likelihood. When $|\psi _\mu\> $ is in the
domain of both $D_\mu ^{1/2}$ and $D_\mu ^{-1/2}$, we can compare the
values of the likelihood as follows. One has 
\begin{equation}
  \mathcal{L}^{sq} =\left( \sum_{\mu}
    \frac{|c_{\mu}|}{\sqrt{\<\psi_{\mu}| D_{\mu} |\psi_{\mu}\>}}
  \right)^2 
\le \left( \sum_{\mu} |c_{\mu}| \sqrt{\<\psi_{\mu}| D_{\mu}^{-1}
      |\psi_{\mu}\>} \right)^2= \mathcal{L}^{opt}~,
\end{equation}
where we used the inequality $\<\psi_{\mu}|
D_{\mu}^{-1}|\psi_{\mu}\>~\<\psi_{\mu}| D_{\mu}|\psi_{\mu}\> \ge 1$,
from Schwartz inequality applied to the vectors $D_{\mu}^{1/2}
|\psi_{\mu}\>$ and $D_{\mu}^{-1/2} |\psi_{\mu}\>$.

\emph{Remark 2: The true value and the most likely one.}  In the
maximum likelihood approach, one optimizes the choice of the POVM in
order to maximize the probability density that the estimated value of
the parameters coincides with the true one. Intuitively, one could
expect that the probability distribution $p(\hat g| g)= \Tr[ M(\hat g)
U_g \rho U_g ^\dag ]$ for the optimal POVM achieves its maximum at the
value $\hat g=g$. Again, this is true for unimodular groups, but fails
to hold for nonunimodular groups.
\begin{prop} 
Let the group $\grp G$ be unimodular. If the covariant POVM $M(\hat
g)$ maximizes the likelihood for a given input state $\rho$, then the
probability distribution $p(\hat g|g)$ of the estimate $\hat g$ on the
state $U_g \rho U_g^{\dag}$ achieves its maximum for $\hat g=g$.
\end{prop} 
\Proof 
Suppose that the most likely value does not coincide with the true
one. Then we can rigidly shift the whole probability distribution with
a post-processing operation that brings the most likely value to the
true one. In fact, if the maximum of $p(\hat g|g)$ occurs at $\hat
g= g h$, we can always replace $M(\hat g)$ with a new covariant POVM $
M'(\hat g)= U_{\hat g } ~\xi '~ U_{\hat g}^{\dag}~$, where
\begin{equation}\label{XiPrime}
\xi'= U_h \xi U_h^{\dag}~.
\end{equation} 
The normalization of the new POVM follows from the fact that for
unimodular groups the DMC operators are trivially proportional to the
identity, and therefore the operator $\xi'$ satisfy the normalization
constraints (\ref{TrXi}) as well.  Moreover, the probability
distribution $p'(\hat g|g)$ associated with $M'(\hat g)$ enjoys the
property $p'(\hat g|g)= p(\hat g h|g)$, whence it achieves the maximum
in $\hat g=g$. In this way, the likelihood of $M'(\hat g)$ would be
higher than the likelihood of the POVM $M(\hat g)$. But this cannot
happen since $M(\hat g)$ is the optimal maximum-likelihood POVM.
Therefore $p(\hat g|g)$ must be maximum in $\hat g=g$.  \qed

For nonunimodular groups the previous argument does not apply, since
the POVM given by (\ref{XiPrime}) is no longer normalized. In fact,
the operator $\xi'$ does not satisfy the normalization constraints
(\ref{TrXi}), since the DMC operators do not commute with the
unitaries $U_h$. In other words, we are not allowed to bring the most
likely value to coincide with the true one by rigidly shifting the
whole probability distribution. As we will see in the explicit example
of the estimation of real squeezing and displacement, this situation
can indeed happen. In order to reduce the discrepancy between the true
value and the most likely one, a suitable choice of the input states
is needed. For example, in the simple case of $\{U_g\}$ being a direct
sum of inequivalent irreps, if the projection of the input state onto
the irreducible subspaces are eigenvectors of the DMC operators, then
the most likely value coincides with the true one. In fact, for any
input state $|\psi\>=\bigoplus_{\mu}~ c_{\mu} |\psi_{\mu}\>$, using
Schwarz inequality, we have
\begin{equation}
  p^{opt}(\hat g|g) = \left| \sum_{\mu} |c_{\mu}| \frac{\<\psi_{\mu}|
      U_{g^{-1} \hat g} D_{\mu}^{-1}
      |\psi_{\mu}\>}{\sqrt{\<\psi_{\mu}| D_{\mu}^{-1}|\psi_{\mu}\>}}
  \right|^2 
\le \left( \sum_{\mu} |c_{\mu}| \sqrt{
      \frac{\<\psi_{\mu}| D_{\mu}^{-2}|\psi_{\mu}\>}{\<\psi_{\mu}|
        D_{\mu}^{-1}|\psi_{\mu}\>}} \right)^2~,
\end{equation}
and if each $|\psi_{\mu}\>$ is eigenvector of $D_{\mu}$, then the last
expression is equal to $p(g|g)$, then the true value is the most
likely one. 
\section{Optimal estimation of real squeezing and displacement}\label{JointEst}
\subsection{Translation and dilation}
In the following we will apply the general framework of Section
\ref{MlEst} to the case of joint estimation of real squeezing and
displacement of a single-mode radiation field with bosonic operators
$a$ and $a^\dag $ with $[a, a^\dag ]=1$. Given the wavefunction of a
pure state $ | \psi \rangle $ in the $X$-representation $\psi (x) =
\langle x |\psi \rangle $, where $|x \rangle $ denotes the
Dirac-normalized eigenstate of the quadrature operator $X=(a +a^\dag
)/2$, the affine transformation on the real line given by $x \to e^{r}
x + x'$ is represented by the unitary transformation
\begin{eqnarray}
\psi (x) \rightarrow e^{-r/2}~\psi (e^{-r}(x-x'))\;,\qquad x',r \in
{\mathbb R} \label{trans}~.
\end{eqnarray}
This transformation corresponds to the action of the unitary operator
$U_{x',r}=D(x')S(r)$ on the ket $|\psi \rangle $, where
\begin{eqnarray}
&&D(x)=\exp\left[x(a^\dag -a)\right ]\;,\nonumber \\ 
&&S(r)=\exp\left[\frac r2 (a^{\dag 2}-a ^2)\right ]\;,
\end{eqnarray}
represent the displacement and the squeezing operator with real
argument, respectively.  In other words, the operators

\begin{equation}
\{U_{x,r}=D(x)S(r)~|~ x, r \in \Reals\}~,
\end{equation} provide a unitary
representation of the affine group in the Hilbert space of
wavefunctions. The affine group is nonunimodular, and in the above parametrization the left- and right-invariant measures are given
 by $\d_L g= e^{-r} \d r \d x$, and $\d_R g= \d r \d x$, respectively.

\subsection{The maximum likelihood POVM}
In order to exploit the general results of Sec. \ref{MlEst}, we need
to know the Clebsch-Gordan decomposition of the representation
$\{U_{x,r}\}$, the irreducible subspaces, and the DMC operators. All
these informations are given in the following, while the proof is
presented in the Appendix.

The Clebsch-Gordan series of the representation $\{U_{x,r}\}$ consists
on two irreps, that we indicate with the symbols $+$ and $-$.
Accordingly, the Hilbert space splits into two irreducible subspaces,
i.e.
\begin{equation}
\spc H= \spc H_+ \oplus \spc H_-~.
\end{equation}
Comparing this decomposition with the general case
(\ref{SpaceDecomp}), we see that the subspaces $\spc H_+$ and $\spc
H_-$ are the representation spaces, while the multiplicity spaces
$\spc M_+$ and $\spc M_-$ are trivially one-dimensional.  The
representation spaces $\spc H_+$ and $\spc H_-$ can be easily
characterized in terms of the quadrature $Y=\frac {a-a^\dag }{2i}$. In
fact, writing the wavefunctions in the $Y$-representation as $\psi
(y)= \<y|\psi\>$, where $|y\>$ are the Dirac-normalized eigenvectors
of $Y$, we have $\spc H_+=\{ |\psi\>~|~ \psi(y)=0 \quad \forall y
<0\}$ and $\spc H_-=\{|\psi\>~|~ \psi (y) =0 \quad \forall y>0\}$.
Therefore, the projection operators onto $\spc H_+$ and $\spc H_-$ can
be written respectively as $\openone_+ = \theta (Y)$ and
$\openone_-=\theta(-Y)$, where $\theta(x)$ is the customary
step-function [$\theta(x)=1$ for $x\ge0$, $\theta(x)=0$ for $x<0$].
Moreover, the DMC operators are
\begin{equation}
D_\pm= \pi ~\frac{\theta (\pm Y)}{|Y|}\;.
\end{equation}
With these tools we are now able to provide the optimal covariant
measurement for the joint estimation of real squeezing and
displacement on a given state of the radiation field. Let us denote by
$|\psi\>$ the input state that undergoes to unknown squeezing and
displacement transformations.  Decomposing the input state on the
subspaces $\spc H_+$ and $\spc H_-$ as $|\psi\>= c_+ |\psi_+\> + c_-
|\psi_-\>$, we can exploit Eq.  (\ref{OptEtaIneq}) and write
explicitly the optimal POVM as $M(x,r)=U_{x,r} |\eta\>\<\eta|
U_{x,r}^{\dag}$ where
\begin{equation}\label{EtaJoint}
|\eta\>= \frac{|Y| \theta(Y)|\psi\>}{\sqrt{\pi~ \<\psi||Y| \theta(Y)
|\psi\>}} + \frac{ |Y| \theta(-Y) |\psi\>}{\sqrt{\pi~\<\psi| |Y|
\theta(-Y) |\psi\>}}~.
\end{equation}
The optimal likelihood is then 
\begin{equation}\label{OptLikJointEst}
\mathcal{L}^{opt}= \frac{1}{\pi}~\left( \sqrt{~\<\psi| ~|Y| \theta(Y) ~|\psi\>} + \sqrt{~\<\psi| ~|Y| \theta(-Y)~|\psi\>} \right)^2~, 
\end{equation}
according to the general expression of Eq. (\ref{OptLikIneq}).
As already mentioned, the expression of the likelihood provides some
insight about the states that are most sensitive in detecting an
unknown combination of real squeezing and displacement. Essentially, one can improve the likelihood by increasing the
expectation value of $|Y|$, the modulus of the quadrature $Y$. In addiction, the use of wavefunctions that in the $Y$-representation
are non-zero both in the positive half-line and in the negative
half-line allows to exploit the interference of the components
$|\psi_+\>$ and $|\psi_-\>$ to enhance the value of the likelihood. 

\subsection{Real squeezing and displacement on a coherent state.}  
Using Eq. (\ref{EtaJoint}) for the optimal POVM, we can obtain the
probability distribution of the estimated squeezing and displacement
parameter for a given input state. In particular, for a coherent input
state $|\alpha\>$ the sensitivity of the measurement can be
significantly improved by increasing the imaginary part of $\alpha$,
this corresponding to taking coherent states with a high expectation
value of $|Y|$. The probability distribution for the joint estimation
of squeezing and displacement on the vacuum state and on a coherent
state with $\alpha =10 i$ has been reported in Figs. 1 and 2,
respectively.  Comparing the two figures, a remarkable improvement in
the precision of the measurement can be observed as an enhancement of
the likelihood, along with a narrowing of the probability
distribution. Moreover, we can observe that for the vacuum state the
maximum of the probability density is not achieved by the true value
(which is given by to $x=r=0$).  The discrepancy between the most likely
value and the true one, due to the fact that the affine group is
nonunimodular, essentially disappears by increasing the expectation
value of $|Y|$. As we can see from Fig. 2, for $\alpha=10 i$ the probability
distribution is approximately a Gaussian centered around the true value
$x=r=0$.

\begin{figure}[htb]
\begin{center}
\includegraphics[scale=1]{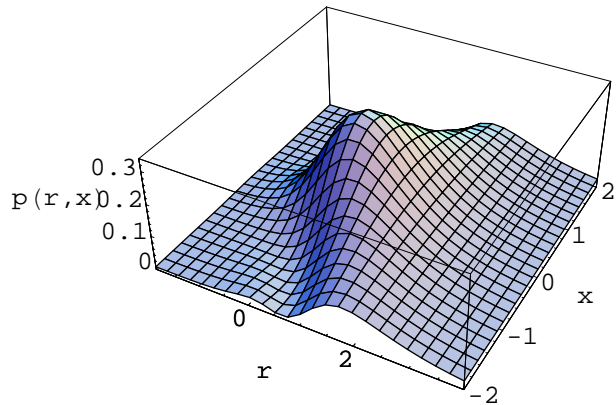}
\caption{Optimal probability distribution for the joint estimation of
  the squeezing parameter and real displacement for the vacuum state.
}
\label{f:fig1}
\end{center}
\end{figure}
\begin{figure}[htb]
\begin{center}
\includegraphics[scale=1]{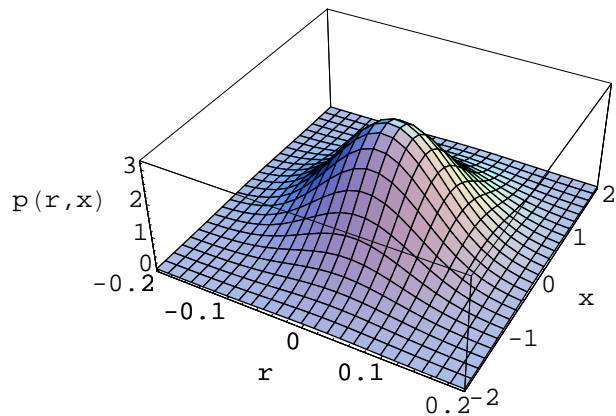}
\caption{Optimal probability distribution for the joint estimation of
  the squeezing parameter and real displacement for a coherent state
  with amplitude $\alpha =10 i$.}
\label{f:fig2}
\end{center}
\end{figure}

Now it is interesting to focus on the asymptotic behavior of the
probability distribution for a coherent state $|i a\>$ with
$a \in \Reals$ going  to infinity. In
the asymptotic regime, the probability distribution
\begin{equation}\label{ProbCoh}
p_{a}(x,r)\d x \d r e^{-r} = \left| \< \eta|~ U_{x,r}~ |i a \> 
\right|^2  \d x \d r e^{-r}~, 
\end{equation}
given by optimal vector $|\eta\>$ in Eq. (\ref{EtaJoint}), can be
further simplified.   In fact, 
the wavefunction of the coherent state can be written as  
\begin{eqnarray}
\<y| ia \> = \left( \frac{2}{\pi} \right)^{1/4} ~ e^{-(y- a)^2}~\;,
\label{wfc}
\end{eqnarray}
and for $a \gg 1$ it is essentially confined in the positive part of
the $y$-axis.  Hence, in the expression (\ref{EtaJoint}) we can
asymptotically neglect the component in the subspace $\spc H_-$ and
drop the modulus from $|Y|$. In this way, the probability distribution
(\ref{ProbCoh}) can be approximated as
\begin{equation}
p_{a}(x,r) \d x \d r e^{-r}  \approx  \frac{1}{\pi} ~
\frac{\left| \<i a | ~U_{x,r} Y~ |ia \> \right|^2}{\<i a|~ Y
  ~| i a \>}~\d x \d r e^{-r}~.
\end{equation}

Neglecting  the higher order terms, we thus obtain the Gaussian distribution
\begin{equation}\label{AsyCoh}
p_{a} (x,r) \d x \d r e^{-r} = \frac{a}{\pi}~ e^{-
  a^2 r^2}~e^{-x^2}~\d x \d r~. 
\end{equation}
Notice that, asymptotically the most likely values of the unknown
parameters $x$ and $r$ are the true ones $x=r=0$, and, in addiction,
also the mean values of $x$ and $r$ coincide with the true one, namely
the estimation is unbiased.  

From the asymptotic expression (\ref{AsyCoh}) can see that the
uncertainty in the estimation of the squeezing parameter $r$ goes to
zero with the number of photons $\bar n= a ^2$, namely the r.m.s
error is $\Delta r= 1/\sqrt{2 \bar n}$, while the uncertainty in the
estimation of the displacement $x$ remains fixed, with the value
$\Delta x= 1/\sqrt{2}$.  It is interesting to compare the precision
achieved by the joint estimation with the precision that could be
achieved if the parameters $x$ and $r$ were measured separately.
First, it is known that the optimal estimation of real displacement is
given by the observable $X= (a+a^{\dag})/2$, and the corresponding
uncertainty in a coherent state is given by $\Delta x^{opt} = 1/2$.
The optimal estimation of real squeezing has been recently derived
in \cite{SqueezEst}, and we report here the asymptotic distribution
for an excited coherent state:
\begin{equation}
p(r)= \sqrt{\frac{2 |\alpha | ^2}{\pi}}~ e^{-2|\alpha|^2 r^2}~. 
\end{equation}
Accordingly, the r.m.s. error for the optimal estimation of squeezing in a coherent state is $\Delta r^{opt} =1/(2 \sqrt{\bar n})$. 
It is remarkable to note the relation
\begin{equation}
\left\{ 
\begin{array}{lll}
\Delta x &=& \sqrt{2}~ \Delta x^{opt} \\
\Delta r &=& \sqrt{2}~ \Delta r^{opt} 
\end{array}
\right.
\end{equation}
between the uncertainties in the joint measurement and the ones in the separate measurements of squeezing and displacement.
In particular, the product of the uncertainties in the joint estimation is twice the product of uncertainties in the optimal separate measurements: 
\begin{equation}\label{UncProd}
\Delta x \Delta r = 2 \Delta x^{opt} ~ \Delta r^{opt} ~.
\end{equation}  
Surprisingly, this is exactly the same relation as the one occurring
in the optimal joint measurement of two conjugated quadratures $X$ and
$Y$ that can be achieved by heterodyne detection \cite{yuen}.

\subsection{Joint estimation for a displaced squeezed state}
In the previous section we analyzed the optimal estimation of
squeezing and displacement for an excited coherent state. In that
case, while the error in the estimation of $r$ goes to zero with the
number of photons, the error in estimating $x$ remains fixed. However,
it is possible to choose the input state in such a way that both
variances vanish in the asymptotic limit. To this purpose, we consider
here displaced squeezed states $|i a ,z\>= D(i a) S(z) |0\>$ with
$a,z \in \Reals$. Such states have the
wavefunction
\begin{equation}\label{DispSqueez}
\<y| i a,z\>= \left( \frac{2 e^{2 z}}{\pi} \right)^{1/4} ~ e^{-(y
  -a)^2 e^{2 z}}~, 
\end{equation}
namely a Gaussian centered around the mean value $a$, with standard
deviation $\sigma = 1/(\sqrt{2} e^{z})$. Clearly, if the conditions
$a \gg 1$ and $a \gg \sigma$ are simultaneously satisfied,
such a Gaussian lies almost completely in the positive half-line.
Therefore, in the asymptotic limit $a \to + \infty$, $ a \gg
e^{-z}$, the optimal probability distribution can be approximated as
\begin{equation}
p_{a,z}(x,r) \approx \frac{1}{\pi}~ \frac{|\<i a, z|~U_{x,r}
Y~|i a,z\>|^{2}}{\<i a,z|~ Y~ |i a,z\>}~,
\end{equation}  
as in the case of coherent states. By calculating the expectation
values and keeping the leading order terms, we then obtain the
asymptotic distribution
\begin{equation}
p_{a,z}(x,r) \d x \d r e^{-r}= 
\frac{a}{\pi} e^{-(a e^z r)^2} e^{-(x e^{-z})^2}\d x \d r~.
\end{equation}
Again, in the asymptotic limit the most likely values in the
probability distribution coincide with the true ones, and, moreover,
the estimation is unbiased.

The r.m.s. error in the estimation of squeezing and displacement are
now given by $\Delta r=1/(\sqrt{2} a e^{z})$ and $\Delta
x=1/(\sqrt{2} e^{-z})$, respectively. In order to have both errors
vanishing, one needs simultaneously $a e^{z}\gg1$ and $e^{-z} \gg 1$.
For example, we can have an isotropic distribution $\Delta r= \Delta
x$ with the choice $a = e^{-2z}$. In the isotropic case, we notice
that only a small fraction of order $\sqrt{\bar n}$ of the total
number of photons $\bar n=a^2 + \sinh^2 z$ comes from squeezing. Since
one has $\Delta x = \Delta r = 1/(\sqrt{2} e^{-z}) \approx 1/(\sqrt{2}
\bar n^{1/4})$, the convergence to the asymptotic regime is quite
slow: the uncertainty goes to zero with order $1/ \bar n^{1/4} $.  As
an example in the asymptotic regime, in Fig. 3 we plot the exact
probability distribution for $\bar n = 4000$.
\begin{figure}[htb]
\begin{center}
\includegraphics[scale=1]{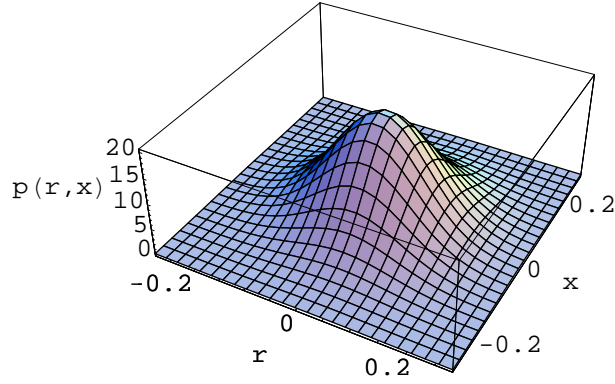}
\caption{Optimal probability distribution for the joint estimation 
  on a displaced squeezed states with total number of photons $\bar n
  =4000$ (with signal photons $\bar n -\sqrt{\bar n}$, and squeezing
  photons $\sqrt{\bar n}$).}
\label{f:fig3}
\end{center}
\end{figure}

Also for displaced squeezed states one can compare the accuracy of the
joint estimation with that of independent measurements of squeezing
and displacement. For squeezed states of the form (\ref{DispSqueez})
the uncertainty in the measurement of the observable $X$ is $\Delta
x^{opt} = 1/(2 e^{-z})$. On the other hand, the optimal estimation of
squeezing is obtained asymptotically by the observable $\ln
(|Y|/a)$, and its uncertainty is $\Delta r^{opt} = 1/(2 a e^z)$ 
(according to the results of \cite{SqueezEst}).  Again, a factor
$\sqrt{2}$ relates the standard deviations of the marginals in the
joint estimation with those of the optimal separate measurements.
Again, we find the relation (\ref{UncProd}) between the product of
uncertainties.

We then see that the estimation of real squeezing and displacement on
input states of the form (\ref{DispSqueez}) corresponds in the
asymptotic regime to the joint measurement of the observables $X$ and
$(\ln |Y|/a)$, whose commutator is $[X, \ln (|Y|/a)]=i/(2
Y)$~. We thus have the Heisenberg-Robertson inequality
\begin{equation}
\Delta X \Delta \ln (|Y|/ a ) \ge \left | ~\<~ \frac{1}{4 Y} ~\>~\right|~,
\end{equation}
where  $\< \ \ \>$ denotes the expectation value.  From this
point of view, the displaced squeezed states (\ref{DispSqueez}) are
characterized asymptotically as minimum uncertainty states, since they
saturate the inequality, and the product of uncertainties in the
maximum likelihood estimation is exactly twice the Heisenberg
limit. To the best of our knowledge, the present case is the first
example of joint measurement of two harmonic-oscillator 
noncommuting observables  whose commutator is not a $c$-number. 
\subsection{Joint estimation of squeezing, displacement, and
  reflection} The results presented in the previous sections can be easily
extended to include the estimation of the reflection on the real line
that is realized by the parity operator $P= (-1)^{a^\dag a}$. In this
case we are interested in estimating the three parameters in the
transformation
\begin{equation}
|\psi\> \longrightarrow P^{\epsilon} D(x) S(r) |\psi\>~,
\end{equation}
where $\epsilon $ can assume the values $0$ or $1$. The representation
$\{P^{\epsilon} D(x) S(r)\}$ is now irreducible in $\spc H$, and the
associated DMC operator is $D= \pi/|Y|$. For a given input state
$|\psi\>$, the optimal POVM can be written as $M(\epsilon,x,r)=
P^{\epsilon} D(x) S(r) |\eta\>\<\eta| S(r)^{\dag} D(x)^{\dag}
P^{\epsilon}$, where
\begin{equation}
|\eta\>=\frac{|Y| |\psi\>}{\sqrt{\pi~\<\psi|~ |Y|~|\psi\>}}~,  
\end{equation}
according to Eq. (\ref{OptEtaIneq}).  Clearly, also in this case it is
possible to enhance the sensitivity of detection by increasing the
average value of the modulus $|Y|$, using coherent states or displaced squeezed states.

\subsection{Perfect detection of squeezing and displacement: a two-mode setup}
A fundamental mechanism leading to the optimal estimation of group
transformations is the use of equivalent representations of the group,
via the technique of entanglement between representation spaces and
multiplicity \cite{DeGiorgi,EntEstimation}.  Such a strategy strongly
improves the quality of estimation for unimodular groups
\cite{CovLik,Refframe,DeGiorgi}, and in the case of compact groups is
a way to obtain the best estimation of a unitary
transformation\cite{EntEstimation}.  Equivalent representations of the
affine group can be obtained by entangling the radiation with a second
reference mode, which is not affected by the unknown squeezing and
displacement.  This corresponds to considering the two-mode Hilbert
space $\spc H \otimes \spc H$ and the representation $\{ U_{x,r}
\otimes \openone \}$ of the affine group, namely the reference mode
plays the role of an infinite dimensional multiplicity space. In this
case, the optimal estimation can be obtained with the POVM specified
by the general formula (\ref{OptEta}). A remarkable feature of the
two-mode setup is that it is possible to have an \emph{orthogonal}
POVM for the estimation, namely there exists an ordinary observable on
the extended Hilbert space, associated with the joint measurement of
squeezing and displacement.   In fact, by defining the bipartite vectors
\begin{equation}\label{Pointer}
|\Phi _{\pm }\>\!\> = \frac{1}{\sqrt \pi}
\int_{-\infty}^{+\infty}~ \d y~ \sqrt{|y|}~|y\>| \pm y\>~,
\end{equation}
we can construct the orthogonal POVM
\begin{equation}\label{ObsAffine}
M(g)=U_{x,r} \otimes \openone  ~ \left (
\sum _{s=\pm } |\Phi _{s}\>\!\>\<\!\<\Phi
_{s}|\right )~ U_{x,r}^{\dag} \otimes \openone ~.
\end{equation}
The normalization follows straightforwardly from Eq. (\ref{Norm}).
Moreover, we have the orthogonality relation
\begin{eqnarray}
&&\<\!\<\Phi _s |~ U_{x,r} \otimes \openone~ |\Phi _{s'} \>\!\>= 
\frac 1\pi \int_{-\infty}^{+\infty} \d y ~|y|~\<y| U_{x,r} |sy/s' \>
\nonumber \\& & 
= \frac{1}{\pi}~ \int_{-\infty}^{+\infty} \d y~ |y|~ e^{-r/2} e^{-2i x y
  e^{-r}}~ \delta (y-ye^{-r})\,\delta_{s,s'}
=  \delta (x) \delta (r) \delta_{s,s'}~,
\end{eqnarray}
where we used Eq. (\ref{dsr}) of the Appendix and the identity $\delta
(y-ye^{-r})= \frac{1}{|y|} \delta (r)$.     The vectors in Eq. 
(\ref{Pointer}) are not normalizable. However, similarly to the case
of the heterodyne operator, where physical states can arbitrarily well
approximate the unnormalizable eigenstates \cite{rc}, here one can
consider, e.g., states of the form 
\begin{equation}
|\Phi (\lambda )_{\pm }\>\!\>= 
N_{\lambda}~ \lambda^{a^{\dag} a +b^{\dag} b}~
|\Phi _\pm \>\!\>~, \end{equation} 
where $0<\lambda<1$, $N_{\lambda}$
is a normalization constant, and $b^{\dag}b$ is the photon number
operator for the auxiliary mode $b$. The states $|\Phi(\lambda )_\pm 
\>\!\>$ approaches the optimal vectors (\ref{Pointer}) as 
$\lambda \to 1$, with a correspondent increasing to 
infinity of the average energy of the radiation field.      

\section{Conclusions}
In this paper we presented the joint estimation of real squeezing and
displacement of the radiation field from the general point of view of
group parameter estimation. The combination of squeezing and
displacement provides a representation of the affine group ``$ax
+b$'', which is the paradigmatic example of a nonunimodular group. To
deal with the concrete example of squeezing and displacement, we
derived in the maximum likelihood approach the optimal estimation of a
group transformation in the case of nonunimodular groups, providing
explicitly the optimal POVM for a given input state. In this analysis,
some remarkable features of estimation showed up. Firstly, while for
unimodular groups the maximum likelihood measurements coincide with
the usual square-root measurements, for nonunimodular groups the SRM
are no longer optimal, namely they do not maximize the probability
density of detecting the correct value. Moreover, for nonunimodular groups one can optimize the estimation strategy in the maximum likelihood approach, but even for the optimal POVM the true value is not the one which is most
likely to be detected. To reduce the discrepancy between the true
value and the most likely one a suitable choice of the input states is
required.  Both these features are in general unavoidable, and their
origin tracks back to the presence of a positive unbounded
operator---the Duflo-Moore-Carey operator---in the orthogonality
relations for nonunimodular groups. In the problem of joint estimating
real squeezing and displacement, all the above effects occur. In
particular, for coherent input states and displaced squeezed states we
observed how an increase in the expectation value of $|Y|$ gives rise to an
improvement in the quality of estimation, along with a reduction of
the discrepancy between the true value and the most likely one.  In the mentioned cases the probability distributions for joint estimation become asymptotically Gaussian, and the r.m.s. errors $\Delta x$ and $\Delta r$ can be easily calculated. Remarkably, the product of uncertainties in the joint estimation is exactly twice the product of uncertainties for the optimal separate measurements of squeezing and displacement, in the same way as in the joint measurement of two conjugated quadratures.   
Finally, the use of entanglement with an additional mode of the
radiation field allows one to perform a von Neumann measurement for
the joint estimation of squeezing and displacement, in terms of an
ordinary observable with continuous spectrum. The ideal input states
for detecting an unknown affine transformation can then be
approximated by normalizable states.  \acknowledgments
G.C. aknowledges M. Hayashi for pointing out
Ref.\cite{OzawaUnpublished}, and ERATO project for hospitality.
M.F.S. acknowledges support from INFM through the project
No. PRA-2002-CLON.  This work has been supported by Ministero Italiano
dell'Universit\`a e della Ricerca (MIUR) through FIRB (bando 2001).

\section{Appendix}
Here we derive the group theoretical structure of the representation
$\{U_{x,r}= D(x)S(r)\}$ of the affine group, by explicitly calculating
the expression for the group average of an operator $A$ over the
left-invariant measure $\d_L g= \d x \d r e^{-r}$.
\begin{prop}
  The Clebsch-Gordan series of $\{U_{x,r}\}$ contains two irreducible
  irreps, $+$ and $-$. Accordingly, the Hilbert space can be
  decomposed as $\spc H= \spc H_+ \oplus \spc H_-$, and the
  projections onto the irreducible subspaces are $\openone_+=\theta
  (Y)$ and $\openone_-=\theta(-Y)$, respectively. The DMC operators
  are given by $D_+=\pi \theta (Y)/|Y|$ and $D_-=\pi \theta(-Y)/|Y|$.
\end{prop}
 
\Proof Using twice the resolution of the identity in terms of the
eigenstates $|y \rangle $ of the quadrature operator $Y$, and the
relation
\begin{eqnarray}
U_{x,r}| y \rangle = D(x)S(r)| y \rangle = 
D(x) e^{-r/2}|e^{-r} y \rangle =
e^{-r/2}e^{-2ixye^{-r}}|e^{-r}y \rangle 
\;,\label{dsr}
\end{eqnarray}
we can calculate the group average as follows 
\begin{eqnarray}
\<A\>&&=\int _{-\infty}^\infty \d r e^{-r}\int _{-\infty}^\infty \d x \, U_{x,r} A U_{x,r}^{\dag}
  \nonumber \\& & =
 \int _{-\infty}^\infty \d r e^{-r}\int _{-\infty}^\infty \d x
\int _{-\infty}^\infty \d y \int _{-\infty}^\infty \d y' \, \langle y|A |y'
\rangle \, e^{-2i x(y-y')e^{-r}}\, e^{-r} |e^{-r} y
\rangle \langle e^{-r} y'| \nonumber \\& & =\pi \int _{-\infty}^\infty \d r
\int _{-\infty}^\infty \d y \, \langle y|A |y \rangle \,e^{-r} |e^{-r} y
\rangle \langle e^{-r} y| \nonumber \\& & =
\pi \int _{-\infty}^\infty \d y
\int _{0}^{\hbox{\scriptsize{sgn}} (y) \infty }\d \tilde r  \,\frac 1 y \,\langle y|A |y
\rangle  \,|\tilde r
\rangle \langle \tilde r| \nonumber \\& &
=\pi \int _{0}^\infty \d \tilde r \, |\tilde r \rangle \langle \tilde r| \, \Tr[A
  \theta (Y)/Y] - \pi
\int _{-\infty}^0 \d\tilde r \,|\tilde r \rangle \langle \tilde r| \,\Tr[A
  \theta (-Y)/Y]\nonumber \\& & 
=\theta (Y) ~\Tr [A~ \pi \theta (Y)/|Y|] + \theta (-Y) ~\Tr [A~ \pi \theta (-Y)/|Y|]
\;.\end{eqnarray}
The thesis follows by comparing the last equation with the general formula (\ref{AveOp}) for the group average.
 
\end{document}